\documentclass[10 pt, letter]{article}
\usepackage{graphicx}
\usepackage{caption}
\usepackage{amsmath}
\usepackage{amsthm}
\usepackage{enumerate}
\usepackage{subfig}
\usepackage{float}
\usepackage{amsfonts}
\DeclareCaptionLabelFormat{Arfken}{\textsc{#1} #2}
\DeclareCaptionLabelSeparator{Arfken}{\quad \ }
\title{Quasi non-Markovian approach to the study of decoherence of a controlled-not quantum gate in a chain of few nuclear spins quantum computer.}
\author{ P. L\'opez\footnote{pablocarloslopez@hotmail.com}~~ and G.V. L\'opez\footnote{gulopez@udgserv.cencar.udg.mx}\hskip1pc \\
\small \emph{Departamento de F\'{i}sica, Universidad de Guadalajara,}\\
\small \emph{Blvd. Marcelino Garc\'{i}a Barragan y Calzada Ol\'{i}mpica, 44200 Guadalajara, Jalisco, Mexico}}
\date{\ \\ \today}

\begin{document}
\maketitle

\begin{abstract}
We develop in the weak coupling approximation a quasi-non-Markovian master equation and study the phenomenon 
of decoherence during the operation of a controlled-not (CNOT) quantum gate in a quantum computer model formed by a
linear chain of three nuclear spins system with second neighbor Ising interaction between them. We compare with the behavior of the Markovian counterpart for temperature different from zero (thermalization) and at zero temperature for low and high dissipation rates.
At low dissipation there is a very small difference between Markovian and quasi no-Markovian at any temperature which is unlikely to be
measured, and at high dissipation there is a difference which is likely to be measured at any temperature. 
\end{abstract}
\centerline{PACS: 03.65.Yz, 03.65 Ud}
\section{Introduction } \label{sec:1}
A quantum open system is generally characterized by a non unitary evolution of the reduced density  matrix associated to of the central system and its interaction with the environment. Different types of approaches have been developed to understand the phenomenon of decoherence that arises in the open quantum systems which it is related to the lost of the interference terms of the product of the quantum wave function\cite{Lo1}-\cite{Plenio1}, that is, the non diagonal elements of the reduced density matrix go to zero value.  Since the complete insulate quantum system is almost impossible to have, 
decoherence becomes an intrinsic phenomenon related to the quantum principles and maybe related to the ``emergent reality`` of the classical world \cite{Zu1}-\cite{Manis2}. Many interest has been created in the phenomenon because of the difficulties it carries to perform quantum computation. Non-Markovian systems or systems where the environment is supposed to keep memory, is a topic in this subject and there is not a unified consensus about the best approach for studying  the dynamics of this systems \cite{Manis1}-\cite{Plenio2}, which  makes non-Markovian to be a very interesting subject. In the Markovian approach the non-unitary evolution equation is called "master equation" which is a differential equation for the traced over the environmental variables of the full density  matrix. In principal, in the non-Markovian approach one will have to obtain an integro-differential equation for the density matrix and to establish the non-Markovian in it, but, how to measure non-Markovian ? it is still uncertain. Every approach needs to be intended to maintain the positiveness and trace equal to one for the reduced density matrix. The best known mathematical formalism which kept these conditions was given by Lindblad~\cite{Lind}, who gave an abstract  general non unitary evolution equation for the reduced density matrix, so keeping the Lindblad form in the equations is a good indication. We thought in quasi non-Markovian as an approximation to a master equation but with a temporal dependence in some of the coefficients which defines the interaction with the environment which also depends on the Ising interaction between the spins and may lead to a different behavior of the traditional Markovian solutions. In addition,  these solutions keep the completely positiveness of the density matrix.\\

We use the weak coupling approximation for a system consisting of a linear chain of three paramagnetic atoms with nuclear spin one half~\cite{Ber1}, interacting with a thermal reservoir (not pure) consisting of a bosonic bath \cite{Sole}-\cite{Oxt}. The temporal dependence  in some terms, in the weak coupling approximation, is what we have considered as something beyond Markovian which is totally related to this type of system, and more specifically,  to the Ising interaction between the nuclear spins.\\

We study the decoherence of quantum controlled-not (CNOT) gates during operation in a quantum computer model. In this work, we are interested in determine the differences between the quasi-non-Markovian and Markovian behavior of a quantum controlled-not (CNOT) gate during its implementation on this model of quantum computer. In the first part of this work, we describe this model and the Hamiltonian of our quantum system interacting with a thermal reservoir, which consist of modes of an electromagnetic field in a cavity where the quantum system is. In the second part,  we perform the weak coupling approximation to obtain a quasi-non-Markovian master equation. We want to point out that, even  this model has not been built, it has been very useful for theoretical studies about implementation of quantum gates and quantum algorithms~\cite{Lo2, Ber2, Lo3} which can be extrapolated to other solid state quantum computers~\cite{Cira1}. Then, the analytical dynamical system of the reduced density matrix elements are obtained, and  the results of the numerical simulations are presented.  We present mainly the differences between  the Markovian and the quasi-non-Markovian behavior on the reduced density matrix elements. 
\section{ Hamiltonian of the system} \label{sec:2}
The Hamiltonian that describes the ideal insulated system of a linear chain of N paramagnetic atoms with nuclear spin one half inside a magnetic field 
\begin{equation} \label{eq:2.01}
 {\bf B}(z,t)=\bigl(b\cos(\omega t+\varphi),b\sin(\omega t+\varphi),B(z)\bigr)\,
\end{equation}
where $b, \omega,$ and $\varphi$ are the amplitude, the angular frequency and the phase of the RF-field, and $B(z)$ represents  the amplitude of the z-component of the magnetic field, is given by~\cite{Lo2}
\begin{equation} \label{eq:2.02}
H_S=-\sum_{k=1}^{N}{\vec\mu}_k\cdot {\bf B}_k+J\sum_{k=1}^{N-1}S_k^zS_{k+1}^z+J'\sum_{k=1}^{N-2} S_k^zS_{k+2}^z\ ,
\end{equation}
where ${\vec\mu}_k$ represent the magnetic moment of the kth-nucleus, which it is given in terms of the nuclear spin as ${\vec\mu}_k=\gamma(S_k^x,S_k^y,S_k^z)$, with $\gamma$ being the proton gyromagnetic ratio and $S_k^j$ being the jth-component of the spin operator, ${\bf B}_k$ represents the magnetic field  Eq. (\ref{eq:2.01}) valuated at the location of the kth-nuclear spin ($z=z_k$). The parameters  $J$ and $J'$ represent the coupling constant at first and second neighbor interaction. The angle between the linear chain and the z-component of the magnetic field is chosen as $\cos\theta=1/\sqrt{3}$ to eliminate the dipole-dipole interaction between the spins.\\�\\
We can write the Hamiltonian (\ref{eq:2.02}) in its diagonal and non diagonal with respect a chosen basis in the $z-$ projection as
\begin{equation}\label{eq:2.02a}
 H_S=H_0+H_{rf}
\end{equation}
where
\begin{equation} \label{eq:2.06}
  H_0=-\sum_{k=1}^{N}\omega_kS_k^z+J\sum_{k=1}^{N-1}S_k^zS_{k+1}^z+J'\sum_{k=1}^{N-2}S_k^zS_{k+2}^z
\end{equation}
and
\begin{equation} \label{eq:2.02b}
  H_{rf}=-\frac{\Omega}{2}\sum_{k=1}^{N}\left(e^{i(\omega t+\varphi)}S_k^{+}+e^{-i(\omega t+\varphi)}S_k^{-}\right)
\end{equation}
Here we have that:  $\omega_k=\gamma B(z_k)$ is the Larmor frequency of the kth-spin, $\Omega=\gamma b$ is the Rabi frequency, and $S_k^{\pm}=S_k^x\pm S_k^y$ represents the ascend operator ($+$) or the descend operator ($-$). The Hamiltonian $H_0$ is diagonal in the basis $\{|\alpha_{N}\dots\alpha_{1}\rangle\}$ with $\alpha_k=0,1$ (one  for the ground state and zero for the exited state ). The action of the spin operators on its respective qubit is given by 
$S_k^z|\alpha_k\rangle=\hbar(-1)^{\alpha_k}|\alpha_k\rangle/2$, 
$S_k^+|\alpha_k\rangle= \hbar \delta_{\alpha_k,1}|0\rangle$, and 
$S_k^{-}|\alpha_k\rangle=\hbar\delta_{\alpha_k,0}|1\rangle$. 
The eigenvalues of $H_0$ in this basis are given by
\begin{equation} \label{eq:2.08}
E_{\alpha_{_{N}}\dots\alpha_1}=-\frac{\hbar}{2}\left\{\sum_{k=1}^{N}(-1)^{\alpha_k}\omega_k+
J\sum_{k=1}^{N-1}(-1)^{\alpha_k+\alpha_{k+1}}+J'\sum_{k=1}^{N-2}(-1)^{\alpha_k+\alpha_{k+2}}\right\} .
\end{equation} 
The elements of this basis forms a register of N-qubits with a total number of $2^N$ registers, which is the dimensionality of our Hilbert space. The allowed transition of one state to another one is gotten by choosing the angular frequency of the RF-field, $\omega$, as the associated angular frequency due to the energy difference of these two levels, and  by choosing the normalized evolution time $\Omega t$ with the proper time duration (so called RF-field pulse). The set of selected pulses defines the quantum gates or the quantum algorithms, and CNOT quantum gate is the gate we want to study.\\�\\
Consider now this system to be immerse in a ''mixed thermal bath of oscillators`` such that the Hamiltonian of the bath is of the form 
\begin{equation}\label{eq:2.09}
H_E=\sum_j^{\infty}\hbar\omega_j a^{\dag}_ja_j. 
\end{equation}
The Hamiltonian of the interaction between the central system and the bath will be taken in the form
\begin{equation}\label{eq:2.10}
 H_I=\sum_{kj}\alpha_{kj}\hat{S}_k\hat{E}_j=\sum_{k,j}^{N,\infty} \left(\alpha^{kj}_{1} S_k^{+}a_{j}+\alpha^{kj}_{2} S_k^{-}a^{\dagger}_{j}\right),
\end{equation}
where the operator $\hat E_j$ is defined as $\hat E_j=a_j+a_j^{\dagger}$,  $S_k$ is the polarization operator, $S_k=S^{+}_k+S^{-}_k$, 
and  we have taken into account  the Jaymes-Cummings rotating wave approximation for the  interaction \cite{JCu},  in order to considerer an excitation-de excitation process of the system trough the coupling with the bath of oscillators with  characteristic frequencies near the resonant frequencies of the transitions. The constants $\alpha^{kj}_{i}$, $i=1...3$ are phenomenological parameters that measures the coupling between the system and the environment and $a_j(a^{\dag}_j)$ are the rising (lowering) operators in jth number of photons in the bath. We can write the total Hamiltonian  as 
\begin{equation} \label{eq:2.11}
H=H_D+W_I\ ,
\end{equation}
where $H_D$ and $W_I$ are given by
\begin{subequations}
\begin{equation}
H_D=H_0+H_E
\end{equation}
and
\begin{equation} \label{eq:2.12}
W_I(t)=H_{rf}+H_I= -\frac{\Omega}{2}\sum_{k=1}^{N}\left(e^{i(\omega t+\varphi)}S_k^{+}+e^{-i(\omega t+\varphi)}S_k^{-}\right)+
\sum_{k,j}^{N,\infty}\left(\alpha^{kj}_1 S_k^{+}a_{j}+\alpha^{kj}_2 S_k^{-}a^{\dag}_{j}\right)\ .
\end{equation}
\end{subequations}
\section{ The weak coupling approximation} \label{sec:3}
Now, for dealing with the non ideal situation we start with the dynamical equation of the evolution of the density matrix for an initially decoupled state in the system plus the environment
\begin{equation}\label{eq:3.01}
\rho=\rho_S\otimes\sigma_E
\end{equation}
where $\rho_S$ is a pure state of the central system and $\sigma_E$ is a thermal stationary mixed state of the environment. In the interaction picture the equation of evolution for the reduced system is   
\begin{equation}\label{eq:3.02}
{d\over dt}\tilde{\rho_S}(t)=
-{i\over \hbar}Tr_E\{[\tilde{W}_I(t),\tilde{\rho}_S(t)\otimes\sigma_E]\},
\end{equation}
where in this interaction picture one has
\begin{equation}\label{eq:3.03}
 \tilde{\rho}_S(t)=e^{iH_0t/\hbar}\rho_Se^{-iH_0t/\hbar}\ ,
 \end{equation}
 \begin{subequations}
 \begin{equation}
\tilde{S}_k^{\pm}(t)=e^{iH_0t/\hbar}S^{\pm}_ke^{-iH_0t/\hbar}=S^{\pm}_ke^{\mp i\Omega_k t},\quad
\end{equation}
and
\begin{equation}\label{eq:3.04}
\tilde{a}_i(t)=e^{iH_Et/\hbar}a_ie^{-iH_Et/\hbar},\quad \tilde{a}_i^{\dag}(t)=e^{iH_Et/\hbar}a_i^{\dag}e^{-iH_Et/\hbar}
\end{equation}
\end{subequations}
with the operator  $\Omega_k$ being defined as
\begin{equation}\label{eq:3.05}
\Omega_k=\omega_k-J(S_{k+1}^z+S_{k-1}^z)-J'(S_{k+2}^z+S_{k-2}^z),
\end{equation}
which commutes with the Hamiltonian $H_0$. The eigenvalues of this operator $\Omega_k$,
\begin{equation}
\Omega_k|i\rangle=\Omega_k^{(i)}|i\rangle\ ,
\end{equation}
  are given in the appendix.\\ \\
The time integration of the system in the interval $[t,t+\Delta t]$ is given as follows
\begin{equation}\label{eq:3.06}
 \tilde{\rho}_S(t+\Delta t)=\tilde{\rho}_S(t)-{i\over \hbar}\int_{t}^{t+\Delta t}dt_1Tr_E\{[\tilde{W}_I(t_1),\tilde{\rho}(t_1)\otimes\sigma_E]\},
\end{equation}
Then by doing a successive change of variables and substituting in (\ref{eq:3.06}), up to second order terms, using Markov approximation and Eq. (\ref{eq:3.01}), we obtain 
\begin{equation}\label{eq:3.07}
\Delta \tilde{\rho}_S={1\over i\hbar}\int_{t}^{t+\Delta t}dt_1Tr_E\{[\tilde{W}_I(t_1),\tilde{\rho}_S(t)\otimes\sigma_E]\}+
\left(1\over i\hbar\right)^2\int_{t}^{t+\Delta t} dt_1\int_t^{t_1}dt_2
Tr_E[\tilde{W}_{I}(t_1),[\tilde{W}_{I}(t_2),\tilde{\rho}_S(t)\otimes\sigma_E]],
\end{equation}
where time locality is shown inside the integration with  the term $\tilde\rho_s(t)$, and we have set $\Delta \tilde{\rho}_S= \tilde{\rho}_S(t+\Delta t)-\tilde{\rho}_S(t)$. 
One would expect that within this weak coupling approximation, the interaction of the central system with the environment would show a perturbation to the closed system. By substituting  the corresponding time dependence form of $\tilde{W}_I(t)$ in (\ref{eq:3.07}),  one can sees that the following relation must be satisfied (notice that $\Omega_k^{(i)}\approx\omega_k$ for all basic state $|i\rangle$)
\begin{equation}\label{eq:3.08} 
|\omega+\Omega_k^{(i)}|\Delta t << 1,\quad\hbox{for i=1,\dots,8}
\end{equation}
which determine the time path length where  
 there is no interaction with a time dependent external field and no interaction between the spins 
 %establish a smaller $\Delta t$ for the path of evolution of the system since the $\Omega_k$, are in the order of the $RF$ frequencies. 
 How smaller this path has to be is not resolved, but definitively not that small compared to the relaxation times of the environment $\tau_E$ such that the Markov approximation still being valid. The lost of the separability of the initial system-environment states $\rho_s\otimes\rho_E$ for a smaller $\Delta t$ could exist since a longer time will have to pass for the evolution in the system and therefore correlations between the system and environment can arise, but in the case when we have a thermal state for the environment which is our case, any correlation generated by the evolution of the central system will rapidly decay. Integrating (\ref{eq:3.07}) and under the condition (\ref{eq:3.08}), the master equation takes the form
\begin{equation}\label{eq:3.09}
{\Delta \tilde{\rho}_S(t)\over \Delta t}={1\over i\hbar}[\tilde{H}_{rf}(t),\tilde{\rho}_S(t)]+
{1\over \Delta t}\left(1\over i\hbar\right)^2\int_{0}^{\Delta t} d\tau\int_t^{t+\Delta t}dt_1
Tr_E[\tilde{H}_{I}(t_1),[\tilde{H}_{I}(t_1-\tau),\tilde{\rho}_S(t)\otimes\sigma_E]],
\end{equation}
where we have made the change of variables $t_2=t_1-\tau$ with $\tau\in (0,\Delta t)$ such that $t_1\in (t+\tau,t+\Delta t)$ and divided all by $\Delta t$. The first term in the right hand side of (\ref{eq:3.09}) describes the ideal part of the dynamics in the interaction picture (von Neuman evolution), and the second part describes the open dynamics.\\ \\
For a thermalized mixed environmental system one can sees that $\langle a(s)a(t)\rangle_E=\langle a^{\dag}(s)a^{\dag}(t)\rangle_E=0,$ then by doing typical calculations consisting in integrating over $t_1$ by using the spectral representation of the $\Omega_k$, performing the wave rotating approximation and regrouping terms,  it follows that 
\begin{equation*} 
{d\tilde{\rho}_S(t)\over dt}={1\over i\hbar}[\tilde{H}_{rf}(t),\tilde{\rho}_S(t)]-
{1\over \hbar^2 }\sum_{k}\int_{0}^{\infty} d\tau\left\{\langle A_k(\tau)A_k^{\dag}\rangle_Ee^{-i\Omega_k\tau}
\left(S_k^{+}S_k^{-}\tilde{\rho}_S(t)-
S_k^{-}\tilde{\rho}_S(t)S_k^{+}\right)\right.
\end{equation*}
\begin{equation}\label{eq:3.10}
\left.+\langle A_k^{\dag}(\tau)A_k\rangle_Ee^{i\Omega_k\tau}
\left(S_k^{-}S_k^{+}\tilde{\rho}_S(t)- 
S_k^{+}\tilde{\rho}_S(t)S_k^{-}\right)+h.c.\right\}
\end{equation}
where the limit $\Delta t\rightarrow dt$ has been taken,  the superior limit in the integrals has been put infinity  since the correlation functions decay exponentially in time, and the following definitions have been made
\begin{equation}\label{eq:3.11}
A_k=\sum_ig_{ik}a_i, \quad A_k^{\dag}=\sum_jg^{*}_{jk}a_j^{\dag}\ .
\end{equation}
The coefficients $g_{ik}$ and $g_{jk}^*$ are related to the coupling of the central system with the environment and depends on the characteristic frequencies of the modes in the neighborhood of each spin. The correlation functions are described by the Fourier transform of certain spectral density, $j(\omega)$,  associated to the continuous  modes in the thermal bath,
\begin{equation}\label{eq:3.11a}
 \langle A(\tau)A^{\dag}\rangle_E=\gamma_o\int_{-\infty}^{\infty}d\omega j(\omega)e^{i\omega\tau}.
\end{equation}
with $\gamma_o=|g|^2$. The correlation functions appearing in the $\alpha$ factor can be written as
\begin{equation}\label{eq:3.12a}
\int_0^{\infty}d\tau e^{ \mp i\Omega_k\tau}\langle A_k(\pm\tau)A_k^{\dag}\rangle_E={1\over 2}\hat{\gamma}_k\pm i\hat{\Gamma}_k,
\quad
\int_0^{\infty}d\tau e^{\pm i\Omega_k\tau}\langle A_k^{\dag}(\pm\tau)A_k\rangle_E={1\over 2}\hat{\gamma}^{\dag}_k\mp i\hat{\Gamma}^{\dag}_k, 
\end{equation}
where we get the operators
\begin{equation}\label{eq:3.13}
\hat{\gamma}_k=2\pi\gamma_oj(\Omega_k),
\quad 
\hat{\gamma}^{\dag}_k=2\pi\gamma_oj^{\dag}(\Omega_k),
\quad \hat{\Gamma}_k=\gamma_o\mathbf{P.V.}\int_{-\infty}^{\infty}{j(\omega)\over \Omega_k-
\omega}d\omega,
\quad \hat{\Gamma}^{\dag}_k=\gamma_o\mathbf{P.V.}\int_{-\infty}^{\infty}{j^{\dag}(\omega)\over \Omega_k-
\omega}d\omega,
\end{equation}
being $\mathbf{P.V.}$ the Cauchy principal value. These operators are diagonal on the above basis,  $\hat{\gamma}_k|i\rangle={\gamma}_k^{(i)}|i\rangle$ for example, and their eigenvalues are denoted with a upper index  (see appendix).\\ \\
By regrouping terms in (\ref{eq:3.10}) and going back to Schr$\ddot{\text{o}}$dingers picture, we obtain the following master equation 
\begin{equation}\label{eq:3.19}
{d\over dt}\rho_S={1\over i\hbar}[H_S+H_L,\rho_S]+\mathcal{L}\rho(t)_1
\end{equation}
with $\mathcal{L}\rho(t)_1$ defined as
%\begin{equation*}
\begin{eqnarray}\label{eq:3.19a}
\mathcal{L}\rho(t)_1&=&{-1\over \hbar^2 }\sum_k\Biggl\{
{\hat\gamma_k\over 2}\left(S_k^{+}S_k^{-}\rho_S-\sum_{n,m}\gamma_k^{(m,n)}(t)
S_k^{-}|n\rangle\rho_S^{(n,m)}\langle m|S_k^{+}\right)\nonumber \\ \nonumber \\
& &+\left(\rho_SS_k^{+}S_k^{-}-\sum_{n,m}\gamma_k^{(m,n)}(t)
S_k^{-}|n\rangle\rho_S^{(n,m)}\langle m|S_k^{+}\right){\hat\gamma_k\over 2}\nonumber \\ \nonumber \\ 
& &+{\hat\gamma^{\dag}_k\over 2}\left(S_k^{-}S_k^{+}\rho_S-\sum_{n,m}\gamma_k^{(n,m)}(t)
S_k^{+}|n\rangle\rho_S^{(n,m)}\langle m|S_k^{-}\right)\nonumber \\ \nonumber \\
& &+\left(\rho_SS_k^{-}S_k^{+}-\sum_{n,m}\gamma_k^{(n,m)}(t)
S_k^{+}|n\rangle\rho_S^{(n,m)}\langle m|S_k^{-}\right){\hat\gamma^{\dag}_k\over 2}\Biggr\}
\end{eqnarray}
where $\rho_S^{(n,m)}=\langle n|\rho_S|m\rangle$ are the matrix elements of the initial reduced density matrix operator, and $H_L$ in Eq. (\ref{eq:3.19}) is given by 
\begin{equation}\label{eq:3.20}
H_L=\hat{\Gamma}_kS_k^{-}S_k^{+}-\hat{\Gamma}^{\dag}_kS_k^{+}S_k^{-},
\end{equation}
which represents a Lamb shift Hamiltonian and can be not considered  in the dynamics since it commutes with the entire  $H_0$ of the central system. In addition,  it only generates a global shift in the spectrum. The time dependent coefficients  are explicitly given by  
\begin{equation}\label{eq:3.20}
\gamma_k^{(m,n)}(t)=e^{i\left(\Omega^{(m)}_k-\Omega^{(n)}_k\right)t},
\end{equation}
and they represent  local phases for the non diagonal terms of the equations of the density matrix. Therefore,  the positiveness and trace equal 1 are still satisfied for the density matrix. These phases depend linearly on the Ising coupling constants and bring about the quasi non-Markovian behavior of the system.\\ \\
If we consider low Ising coupling with respect the Larmor frequencies, then we can make the following approximation 
\begin{equation}\label{eq:3.21}
|\Omega_k^{(n)}|\approx |\omega_k|,
\end{equation}
and (\ref{eq:3.19}) takes the following form
\begin{equation}\label{eq:3.22}
{d\over dt}\rho_S={1\over i\hbar}[H_S,\rho_S]+\mathcal{L}\rho_2
\end{equation}
where the term $\mathcal{L}\rho_2$ is given by
\begin{equation}\label{eq:3.23}
\mathcal{L}\rho_2={-1\over \hbar^2 }\sum_k\left\{
{\gamma_k\over 2}\left(S_k^{+}S_k^{-}\rho_S-2S_k^{-}\rho_SS_k^{+}+
\rho_SS_k^{+}S_k^{-}\right)+\right.\left.{\gamma^{\dag}_k\over 2}\left(S_k^{-}S_k^{+}\rho_S-2S_k^{+}\rho_SS_k^{-}+
\rho_SS_k^{-}S_k^{+}\right)\right\}, 
\end{equation}
with the coefficients $\gamma_k$ and $\gamma_k^{\dagger}$ written as
\begin{equation}\label{eq:3.23a}
\gamma_k=2\pi\gamma_oj(\omega_k),\quad \gamma^{\dag}_k=2\pi\gamma_oj^{\dag}(\omega_k). 
\end{equation}
This type of master equations generates  no-correlated thermalized cases which describes spontaneous and thermally induced emission-absorption process \cite{BrePe}, \cite{Lo5}, \cite{Sumanta}. The environment generates excitations or de-excitations in the closed system by absorbing-emitting photons of the thermal bath. In this work, we want to establish the differences between Eq. (\ref{eq:3.19}) which may describe a quasi-non-Markovian process via the oscillating term in the non diagonal elements of the dissipator,  and Eq. (\ref{eq:3.22}), which is the typical Markovian master equation for a system immerse in a bosonic field.
\section{Physical quantities}\label{sec:4}
Let us considered a thermal bath of radiation modes at a temperature $T$. The environmental density matrix is given by
\begin{equation}\label{eq:4.04}
\sigma_E={1\over Z_E}e^{-\beta H_E}={1\over\prod_j \sum_n e^{-\beta \omega_j\hbar n}}e^{-\beta\sum_j\omega_j\hbar 
a_{j}^{\dag}a_{j}}
=\prod_{j}\left(1-e^{-\beta\omega_j\hbar}\right)e^{-\beta\omega_j\hbar a_{j}^{\dag}a_{j}}.
\end{equation}
The interaction Hamiltonian between the central system and the environment is represented by a coupling between the polarization operator and a bosonic modes operators. The correlation functions involved in the system ($\langle A_k(\pm\tau)A^{\dag}_k\rangle_E$, $\langle A_k^{\dag}(\pm\tau)A_k\rangle_E$) are calculated,
\begin{equation}\label{eq:4.06}
 \langle A_k(\pm\tau)A^{\dag}_k\rangle_E=
\sum_j|g|^2_{jk}e^{\mp i\omega_j\tau}\left({1\over e^{\beta\omega_j\hbar}-1}+1\right),\quad
\langle A_k^{\dag}(\pm\tau)A_k\rangle_E=
\sum_j|g|^2_{jk}{e^{\pm i\omega_j\tau}\over e^{\beta\omega_j\hbar}-1}.
\end{equation}
The sum over $i$ is dense (there are an uncountable number of radiation modes). If the volume containing this modes is large enough, we can go from a discrete distribution to a continuous distribution of the characteristic frequencies of the radiation modes. The number of characteristic frequencies with wave vector components $\vec{f}$ in the interval $df_xdf_ydf_z$ in the volume $V$ is given by 
\begin{equation}\label{eq:4.07}
{V\over (2\pi)^3}4\pi f^2df={V\omega^2\over\pi^2 c^3}d\omega,
\end{equation}
 where $f=c\cdot \omega$.  Thus the sum in the correlation functions can be changed by an integration over de frequencies with the proper weight factor,
\begin{equation}\label{eq:4.08}
 \langle A_k(\pm\tau)A^{\dag}_k\rangle_E=
{V|g|^2_{k}\over\pi^2 c^3}\int_{-\infty}^{\infty}
d\omega \omega^3 (N(\omega)+1)e^{\mp i\omega\tau},\quad
 \langle A_k^{\dag}(\pm\tau)A_k\rangle_E=
{V|g|^2_{k}\over\pi^2 c^3}\int_{-\infty}^{\infty}d\omega \omega^3 N(\omega)
e^{\pm i\omega\tau}
\end{equation}
where  we have taken a linearly dependence on the characteristic frequencies of the radiation modes, $|g|_{jk}^2=|g_k|^2\omega_j$, and  the Planck's distribution function,
\begin{equation}\label{eq:4.08a}
 N(\omega)={1\over e^{\beta\omega\hbar}-1}\ .
\end{equation}
Comparing this results with the definitions in (\ref{eq:3.12a}) and (\ref{eq:3.13}), one can sees that
\begin{equation}\label{eq:4.09}
j(\Omega_k)=\Omega_k^3 \left(N(\Omega_k)+1\right),
\quad 
j^{\dag}(\Omega_k)=\Omega_k^3 N(\Omega_k),
\end{equation}
and
\begin{equation}\label{eq:4.11}
\gamma_o={V|g|^2_{k}\over\pi^2 c^3}.
\end{equation}
Once we get the definitions of all these constants, we can proceed to solve the above equations. The evolution equations of the matrix elements  are given in appendix.
\section{Simulations and results}\label{sec:5}
Our registers are made up of three qubits $|ABC\rangle$ with $A,B,C=0,1$, or written them with decimal notation, $|1\rangle=|000\rangle$, $ |2\rangle=|001\rangle$ and so on. The parameters used for our simulation are taken from  \cite{Lo3} regarding the Larmor frequencies of the nuclear spins and we take a higher Ising coupling strength for modelling the differences between the Markovian and the quasi-non Markovian regime but maintaining the 2$\pi$ k method \cite{Lo3}. These parameters are (in units of $2\pi~MHz$) as
\begin{equation}\label{eq:5.01}
 \omega_A=400\ ,\quad\omega_B=200\ ,\quad\omega_C=100\ ,\quad J=25\ ,\quad\hbox{and}\quad J'=1
 \end{equation}
There is still one free parameter which is the strength of the coupling between the environment and the central system $|g|^2_k$. This will allow us to model high or low dissipation rates of an homogeneous or inhomogeneous environments. We take the assumption that the environment is acting homogeneously on each qubit, that  is, there is a set of baths of  characteristic frequencies  affecting more closely the resonant frequencies of each spin.\\ \\ 
The  reduced density matrix is then made up of $8\times 8$ complex elements, and if the initial state  is always taken as the exited state $|1\rangle=|000\rangle$, this means that the initial reduced density matrix has the values $\rho_{11}=1$ and $\rho_{ij}=0$ for $i,j\not=1$.
\subsubsection{Controlled-Not (CNOT) quantum gate}  
To get the CNOT quantum gate starting from the ground state $|1\rangle=|000\rangle$, one applies a $\pi/2$- pulse between this state and the state $|3\rangle=|010\rangle$, with resonant frequency $\omega=\omega_B-J$, to get the superposition state $(|1\rangle+|3\rangle)/\sqrt{2}$. Then,  one applies a resonant $\pi$-pulse between the states $|3\rangle$ and $|4\rangle=|011\rangle$, with resonant frequency $\omega=\omega_C+J/2-J'/2$, to get the final desired state $(|1\rangle+|4\rangle)/\sqrt{2}$ which means that  the expected CNOT density matrix would be such that $\rho_{11}=\rho_{14}=\rho_{41}=\rho_{44}=1/2$, and all the other elements are equal to zero. In addition, one allows the system to have two and a half more resonant $\pi$-pulses to have a better look of the CNOT behavior.\\ \\
\subsubsection{Dynamics at room temperature}
We start modeling the dynamics by considering that the environment is  at room temperature ($T=300$ Kelvins). This assumption will make the system to evolve into a thermalized mixed state. We present in the following figures the differences of the behavior of the diagonal terms and the coherent terms involved in the CNOT quantum gate.\\
Figure 1 shows the behavior of the diagonal elements of the CNOT gate for low ($\gamma\approx J'\times 10^{-3}$)  and high ($\gamma\approx J'\times 10^{-1}$) dissipation rates  (the high dissipation rate  is still in the limit of considering the approximation of a perturbation of the central system) for the Markovian and semi(quasi)-non Markovian regimes at a temperature $T=300K$. We can see that the differences between them is very small but still distinguish. For each case, Markovian and quasi-non Markovian, the environment will lead the central system into a thermalized mixture states, with a thermalization time depending on the coupling constant with the environment. We need to point out that this difference increases as the spin coupling constant a first neighbor increases its value.  

\begin{figure}[H] \label{fig:diagt3}
\includegraphics[width=0.8\textwidth, angle=-90]{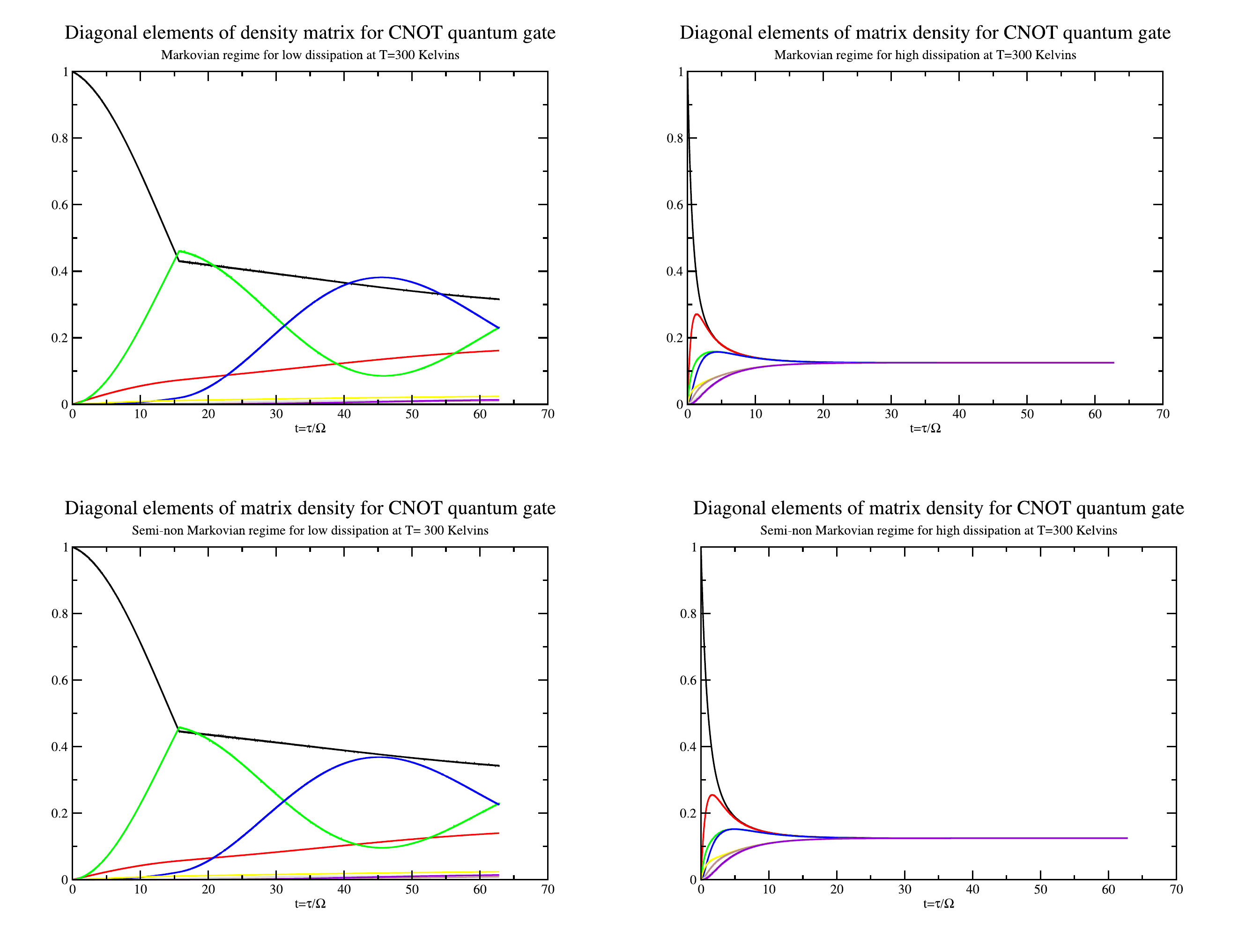}
\centering
    \caption{Diagonal elements of the density matrix for the CNOT quantum gate for low (left) and high (right) rates of 
dissipation in the Markovian and quasi-non Markovian regime at T=300 Kelvins.} 
\end{figure}

Figure 2 shows the behavior of the coherent terms involved in the CNOT gate. We can see that for high dissipation rates a fast thermalization of the system, making the coherent terms goes to zero very rapidly. The term ($|\rho|_{13}$) is related to the first pulse which makes the superposition state for  the CNOT formation. Therefore,  it has higher amplitude, since it will take some time for the environment to completely thermalize the whole system. For the last $\pi$-pulse for the  CNOT formation,   we see that the decoherence is already high. There is a very small sudden birth of coherence in the term $|\rho|_{14}$, due to the pulses of the magnetic field needed to perform the quantum gate. For the low dissipation cases ($\gamma\approx J'\times 10^{-1}$), the quasi-non Markovian and the Markovian regime are very similar, yet we can see a very small differences in the amplitude of the coherent terms. The elements involving  the higher energy level ( $|1\rangle=|000\rangle$)  their amplitude seem to have a grater amplitude for the quasi-non Markovian regime than in the Markovian regime. This situation is contrary in the element $|\rho_{34}|$.  
\begin{figure}[H] \label{fig:coh1314t3}
\includegraphics[width=1.1\textwidth, angle=-90]{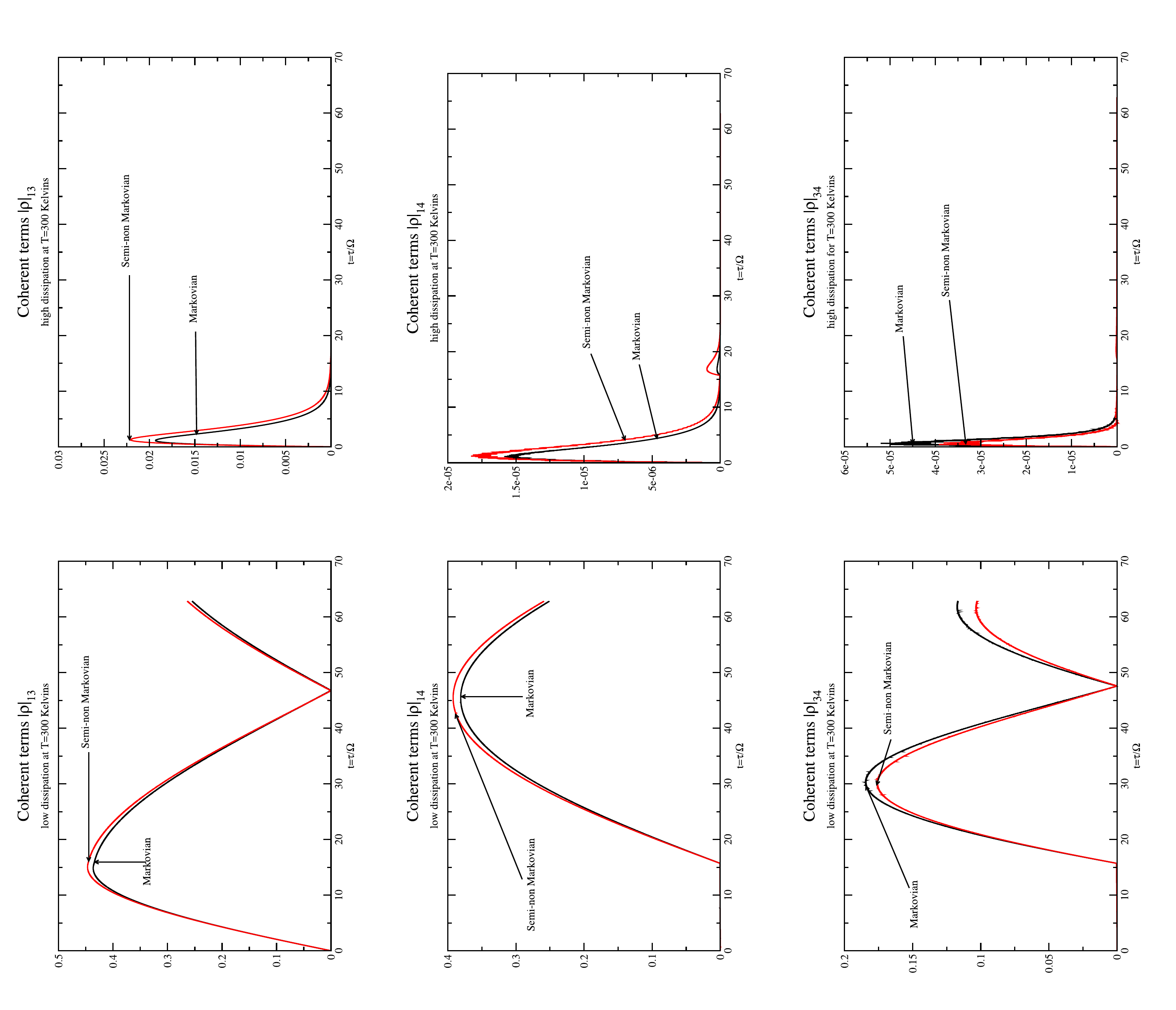}
\centering
    \caption{Coherent elements $|\rho_{13}|$, $|\rho_{14}|$ and $|\rho_{34}|$ of the density matrix for the CNOT quantum gate for low (left) and high (right) rates of 
dissipation in the Markovian and quasi-non Markovian regime at T=300 Kelvins.} 
\end{figure}
\subsubsection{Dynamics at the thermal vacuum}
At $T=0$ Kelvins, the master equation takes the form as described in \cite{Lo5} for the A-Independent environment cases, and we still have the time dependent terms on the non diagonal elements of the master equation, referring to the quasi-non Markovian case. Figure 3 shows the behavior of the diagonal elements of the CNOT quantum gate at $T=0$ Kelvins. We can not see a distinguished difference between the Markovian and quasi-non Markovian regimes as we did  in Figure 1. However,  at high dissipation rate, we still seen for both cases (Markovian and quasi-non Markovian), the rise of the equilibrium ground state at the end of the whole process. This happens because our initial state  is the most exited state, and during the process of dissipation, the environment is not giving off any energy to the system, the quantum system will deliver the energy to all the other states, exiting them. Therefore,  by dissipation,  all of them go back to zero, leaving the system in the ground state $|8\rangle=|111\rangle$ (purple curve in the figures).  
\begin{figure}[H] \label{fig:diagt0}
\includegraphics[width=0.75\textwidth, angle=-90]{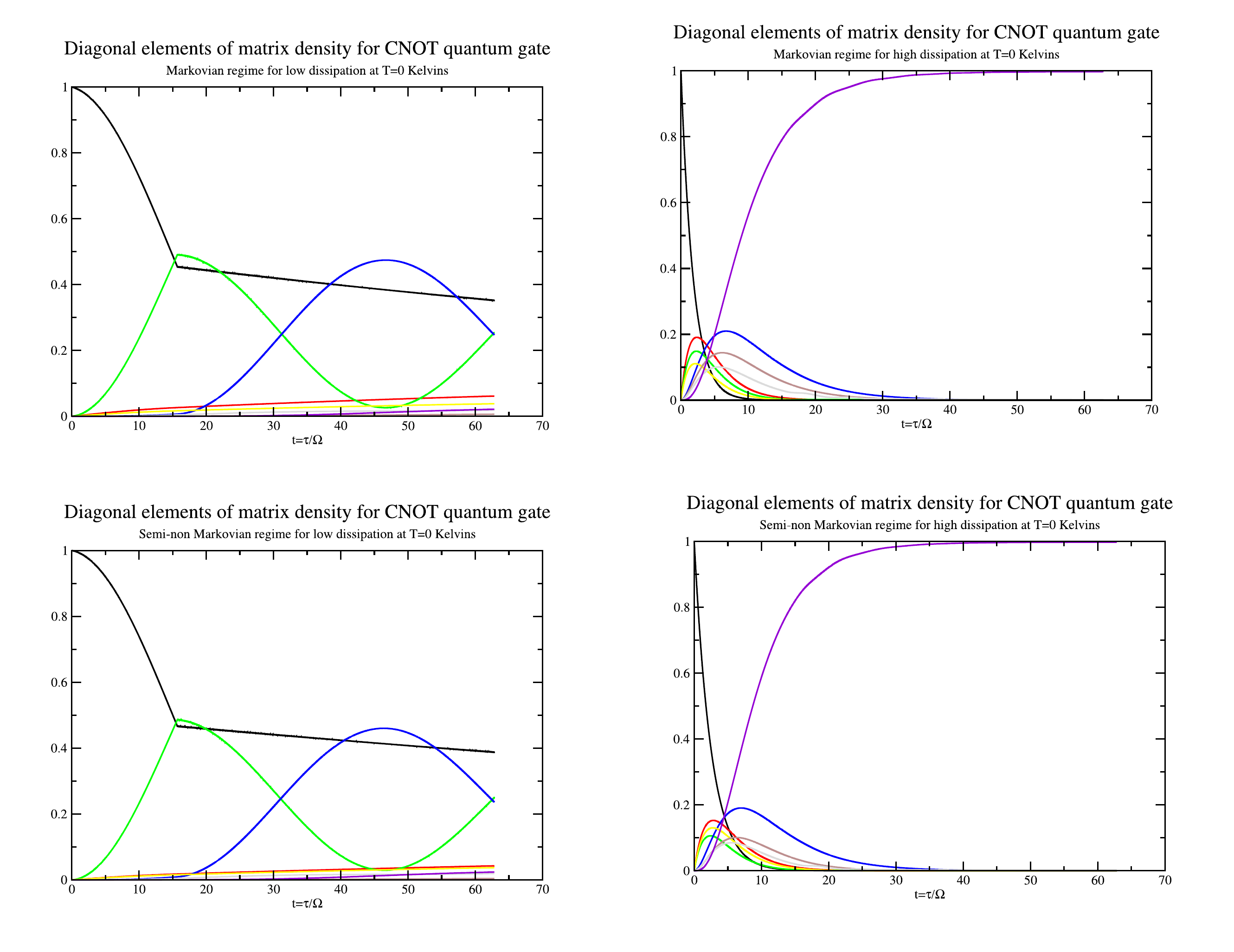}
\centering
    \caption{Diagonal elements of the density matrix for the CNOT quantum gate for low (left) and high (right) rates of 
dissipation in the Markovian and quasi-non Markovian regime at T=0 Kelvins.} 
\end{figure}
Figure 4 shows the behavior of the coherent terms of the reduced density matrix at $T=0^oK$.  For low dissipation rates ($\gamma\approx J'\times 10^{-3}$),
we see a similar behavior as in the room temperature cases.
For high dissipation rates ($\gamma\approx J'\times 10^{-1}$), we see more significant differences between the Markovian and the quasi-non Markovian cases, suggesting that this effect could be observable experimentally.
\begin{figure}[H] \label{fig:coh1314t0}
\includegraphics[width=1.0\textwidth, angle=-90]{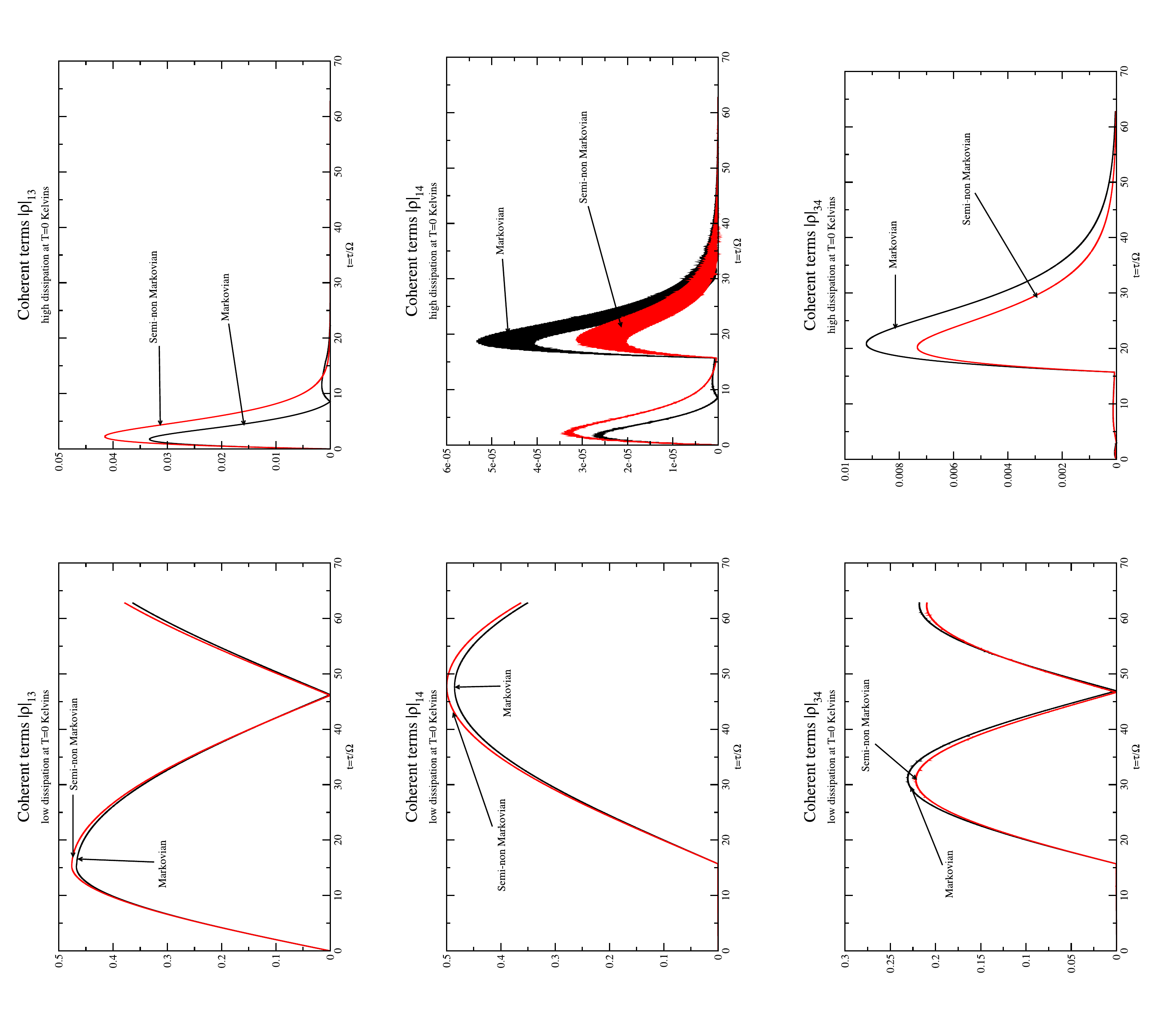}
\centering
    \caption{Coherent elements $|\rho_{13}|$, $|\rho_{14}|$ and $|\rho_{34}|$ of the reduced density matrix for 
 low (left)  and high (right)  rates of 
dissipation in the Markovian and quasi-non Markovian regime at $T=0^o K$.} 
\end{figure}
\subsubsection{Purity calculations.}
The purity function, $P(t)=tr(\rho^2)$, is a measure of how close  a quantum system is from its description as
a pure  state quantum system (the density matrix be written in term of a wave function $\rho=|\Psi\rangle\langle\Psi|$) and varies between 1 and 1/d (d the dimensionality of the density matrix).
This function may decay with the decoherence since the system may move away from an initial pure state.  Therefore, this function can be used to characterize the environment. 

Figure 5 shows the behavior of the purity for the  
CNOT gate at room temperature and at $T=0^oK$. At room temperature it is observed a thermalization of the system, and at $T=0^oK$ a recovery of the purity since the system goes to the quantum ground state, depending on the dissipation rate.

\begin{figure}[H] \label{fig:pur}
\includegraphics[width=0.8\textwidth, angle=-90]{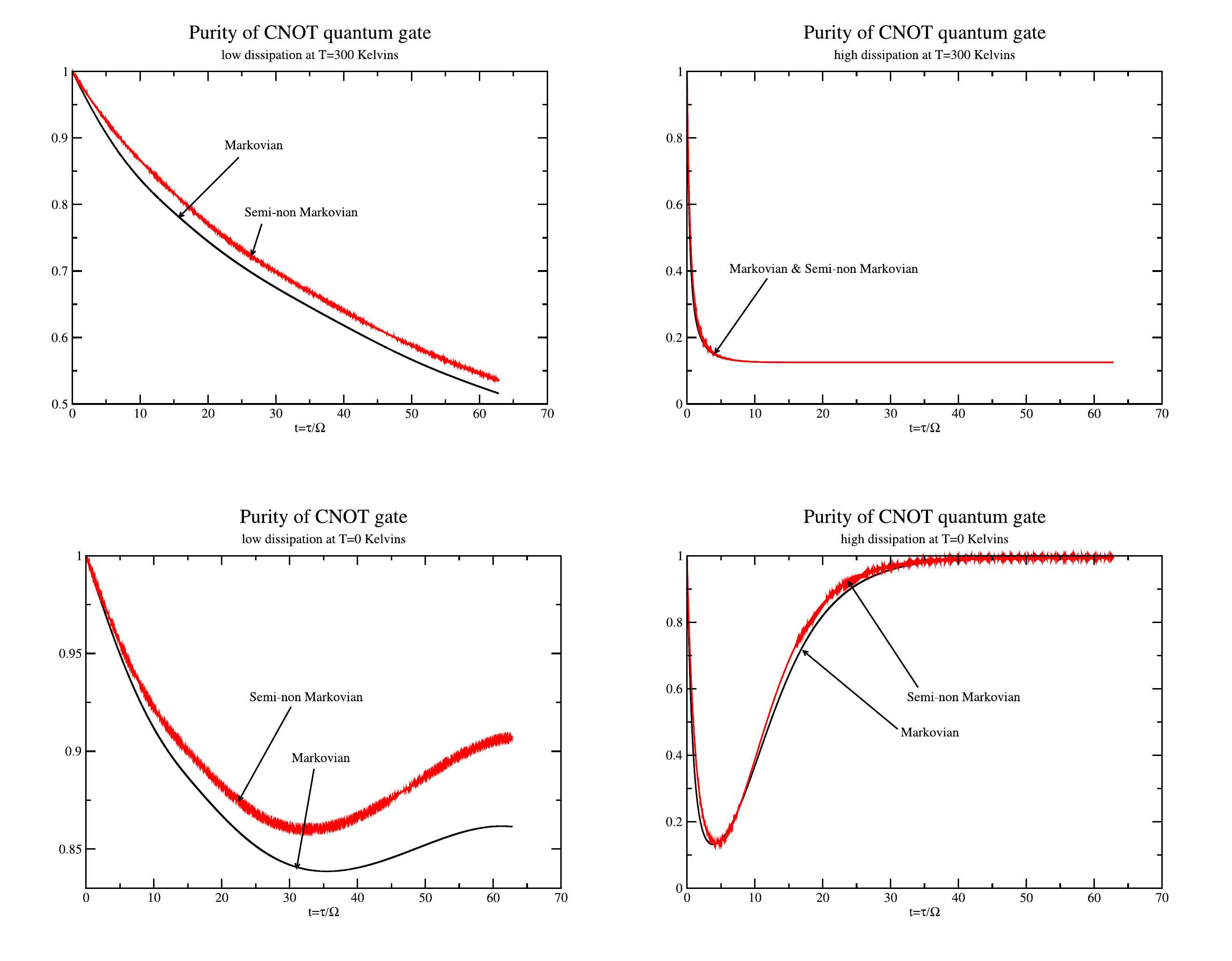}
\centering
    \caption{Purity for the Markovian and quasi-non Markovian regimes at T=300 and T=0 Kelvins for low (left)  and hight (right) dissipation rates.} 
\end{figure}
\section{Conclusions}\label{sec:6}
Within the weak coupling approximation for the study of quantum discrete system with environment, we have obtained a quantum master equation with a time dependent non diagonal dissipative coefficients which shows a quasi-non Markovian behavior. We have solved numerically the master equation for the reduced density matrix associated to our linear chain of three nuclear spin system interacting with the environment. We have made the simulation of CNOT quantum gate operating in this dissipative environment and within the validity of this approximation. We have study the behavior of system-environment interaction with this  quasi-non Markovian master equation and compared the results with the Markovian counterpart. The decoherence of this quantum logic gate have been determined, and  we have seen a different behavior of  the decoherence with the quasi-non Markovian and with  Markovian master equations. 

This difference between  quasi-non Markovian  and Markovian approaches on the reduced density matrix elements grows with the dissipation coefficients  defined in the master equations, but  the diagonal elements remain almost identical over the two types of process. Therefore, for low dissipation  the measuring apparatus will not bring any information of the environmental interaction for a Markovian or quasi-non Markovian process, and for high dissipation this difference can, in principle, be measure. In addition, this difference must increase as the spin coupling parameter a fist neighbor increases  since the spectrum becomes  much more well defined.  This comparison was also made using the purity parameter.  For strong dissipation at $T=0$ Kelvins, we found that purity may increase because, the condition  $tr\rho=1$ on the density matrix, and this implies excitation of the equilibrium state involved in the dynamics (ground state), causing the system to try to return to a pure quantum state description.     
\newpage
\section{Appendix }\label{sec:7}
The evolution equation for the density matrix elements are given from Eq. (\ref{eq:3.19}) by
$$
 \frac{d\rho_{\alpha\beta}}{dt}=-\frac{i}{\hbar}[H,\rho]_{\alpha\beta}+\biggl[{\cal L}(\rho)\biggr]_{\alpha\beta}\
 ,\quad \alpha,\beta=1,\dots, 8.\eqno(X1)$$   
Making the following definition
$$
(vN)_{\alpha\beta}=\frac{1}{i\hbar}[H,\rho]_{\alpha\beta}\quad\quad\hbox{and}\quad 
\mathcal{L}\rho_{_{\alpha\beta}}=\frac{1}{i\hbar}\biggl[{\cal L}(\rho)\biggr]_{\alpha\beta}\ ,\eqno(X2)$$
one gets
\subsection*{Von Neuman (vN) part.}
$$(vN)_{11}=-{\Omega\over 2}e^{i(\omega t+\varphi)}\left(\rho_{21}+\rho_{31}+\rho_{51}\right)
+{\Omega\over 2}e^{-i(\omega t+\varphi)}\left(\rho_{12}+\rho_{13}+\rho_{15}\right)\eqno(V1)$$
$$(vN)_{12}=-\left(\omega_C-j/2-j^{'}/2\right)\rho_{12}-{\Omega\over 2}e^{i(\omega t+\varphi)}\left(\rho_{22}+\rho_{32}+\rho_{52}-\rho_{11}\right)
+{\Omega\over 2}e^{-i(\omega t+\varphi)}\left(\rho_{14}+\rho_{16}\right)\eqno(V2)$$
$$(vN)_{13}=-\left(\omega_B-j\right)\rho_{13}-{\Omega\over 2}e^{i(\omega t+\varphi)}\left(\rho_{23}+\rho_{33}+\rho_{53}-\rho_{11}\right)
+{\Omega\over 2}e^{-i(\omega t+\varphi)}\left(\rho_{14}+\rho_{17}\right)\eqno(V3)$$
$$(vN)_{14}=-\left(\omega_B+\omega_C-j/2-j^{'}/2\right)\rho_{14}-{\Omega\over 2}e^{i(\omega t+\varphi)}\left(\rho_{24}+\rho_{34}+\rho_{54}
-\rho_{12}-\rho_{13}\right)
+{\Omega\over 2}e^{-i(\omega t+\varphi)}\rho_{18}\eqno(V4)$$
$$(vN)_{15}=-\left(\omega_A-j/2-j^{'}/2\right)\rho_{15}-{\Omega\over 2}e^{i(\omega t+\varphi)}\left(\rho_{25}+\rho_{35}+\rho_{55}-\rho_{11}\right)
+{\Omega\over 2}e^{-i(\omega t+\varphi)}\left(\rho_{17}+\rho_{16}\right)\eqno(V5)$$
$$(vN)_{16}=-\left(\omega_A+\omega_C-j\right)\rho_{16}-{\Omega\over 2}e^{i(\omega t+\varphi)}
\left(\rho_{26}+\rho_{36}+\rho_{56}-\rho_{12}-\rho_{15}\right)
+{\Omega\over 2}e^{-i(\omega t+\varphi)}\rho_{18}\eqno(V6)$$
$$(vN)_{17}=-\left(\omega_A+\omega_B-j/2-j^{'}/2\right)\rho_{17}-{\Omega\over 2}e^{i(\omega t+\varphi)}
\left(\rho_{27}+\rho_{37}+\rho_{57}-\rho_{13}-\rho_{15}\right)
+{\Omega\over 2}e^{-i(\omega t+\varphi)}\rho_{18}\eqno(V7)$$
$$(vN)_{18}=-\left(\omega_A+\omega_B+\omega_C\right)\rho_{18}-{\Omega\over 2}e^{i(\omega t+\varphi)}
\left(\rho_{28}+\rho_{38}+\rho_{58}-\rho_{14}-\rho_{16}-\rho_{17}\right)\eqno(V8)$$
$$(vN)_{22}=-{\Omega\over 2}e^{i(\omega t+\varphi)}\left(\rho_{42}+\rho_{62}-\rho_{21}\right)
+{\Omega\over 2}e^{-i(\omega t+\varphi)}\left(-\rho_{12}+\rho_{26}+\rho_{24}\right)\eqno(V9)$$
$$(vN)_{23}=-\left(\omega_B-\omega_C-j/2+j^{'}/2\right)\rho_{23}-{\Omega\over 2}e^{i(\omega t+\varphi)}
\left(\rho_{43}+\rho_{63}-\rho_{21}\right)
+{\Omega\over 2}e^{-i(\omega t+\varphi)}\left(-\rho_{13}+\rho_{24}+\rho_{27}\right)\eqno(V10)$$
$$(vN)_{24}=-\omega_B\rho_{24}-{\Omega\over 2}e^{i(\omega t+\varphi)}
\left(\rho_{44}+\rho_{64}-\rho_{22}-\rho_{23}\right)
+{\Omega\over 2}e^{-i(\omega t+\varphi)}\left(-\rho_{14}+\rho_{28}\right)\eqno(V11)$$
$$(vN)_{25}=-\left(\omega_A-\omega_C\right)\rho_{25}-{\Omega\over 2}e^{i(\omega t+\varphi)}
\left(\rho_{45}+\rho_{65}-\rho_{21}\right)
+{\Omega\over 2}e^{-i(\omega t+\varphi)}\left(-\rho_{15}+\rho_{27}+\rho_{26}\right)\eqno(V12)$$
$$(vN)_{26}=-\left(\omega_A-j/2+j^{'}/2\right)\rho_{26}-{\Omega\over 2}e^{i(\omega t+\varphi)}
\left(\rho_{46}+\rho_{66}-\rho_{22}-\rho_{25}\right)
+{\Omega\over 2}e^{-i(\omega t+\varphi)}\left(-\rho_{16}+\rho_{28}\right)\eqno(V13)$$
$$(vN)_{27}=-\left(\omega_A+\omega_B-\omega_C\right)\rho_{27}-{\Omega\over 2}e^{i(\omega t+\varphi)}
\left(\rho_{47}+\rho_{67}-\rho_{23}-\rho_{25}\right)
+{\Omega\over 2}e^{-i(\omega t+\varphi)}\left(-\rho_{17}+\rho_{28}\right)\eqno(V14)$$
$$(vN)_{28}=-\left(\omega_A+\omega_B+j/2+j^{'}/2\right)\rho_{28}-{\Omega\over 2}e^{i(\omega t+\varphi)}
\left(\rho_{48}+\rho_{68}-\rho_{24}-\rho_{26}-\rho_{27}\right)
+{\Omega\over 2}e^{-i(\omega t+\varphi)}\left(-\rho_{18}\right)\eqno(V15)$$
$$(vN)_{33}=-{\Omega\over 2}e^{i(\omega t+\varphi)}\left(\rho_{43}+\rho_{73}-\rho_{31}\right)
+{\Omega\over 2}e^{-i(\omega t+\varphi)}\left(-\rho_{13}+\rho_{34}+\rho_{37}\right)\eqno(V16)$$
$$(vN)_{34}=-\left(\omega_C+j/2-j^{'}/2\right)\rho_{34}-{\Omega\over 2}e^{i(\omega t+\varphi)}
\left(\rho_{44}+\rho_{74}-\rho_{32}-\rho_{33}\right)
+{\Omega\over 2}e^{-i(\omega t+\varphi)}\left(-\rho_{14}+\rho_{38}\right)\eqno(V17)$$
$$(vN)_{35}=-\left(\omega_A-\omega_B+j/2-j^{'}/2\right)\rho_{35}-{\Omega\over 2}e^{i(\omega t+\varphi)}
\left(\rho_{45}+\rho_{75}-\rho_{31}\right)
+{\Omega\over 2}e^{-i(\omega t+\varphi)}\left(-\rho_{15}+\rho_{37}+\rho_{36}\right)\eqno(V18)$$
$$(vN)_{36}=-\left(\omega_A-\omega_B+\omega_C\right)\rho_{36}-{\Omega\over 2}e^{i(\omega t+\varphi)}
\left(\rho_{46}+\rho_{76}-\rho_{32}-\rho_{35}\right)
+{\Omega\over 2}e^{-i(\omega t+\varphi)}\left(-\rho_{16}+\rho_{38}\right)\eqno(V19)$$
$$(vN)_{37}=-\left(\omega_A+j/2-j^{'}/2\right)\rho_{37}-{\Omega\over 2}e^{i(\omega t+\varphi)}
\left(\rho_{47}+\rho_{77}-\rho_{33}-\rho_{35}\right)
+{\Omega\over 2}e^{-i(\omega t+\varphi)}\left(-\rho_{17}+\rho_{38}\right)\eqno(V20)$$
$$(vN)_{38}=-\left(\omega_A+\omega_C+j\right)\rho_{38}-{\Omega\over 2}e^{i(\omega t+\varphi)}
\left(\rho_{48}+\rho_{78}-\rho_{34}-\rho_{36}-\rho_{37}\right)
+{\Omega\over 2}e^{-i(\omega t+\varphi)}\left(-\rho_{18}\right)\eqno(V21)$$
$$(vN)_{44}=-{\Omega\over 2}e^{i(\omega t+\varphi)}\left(\rho_{84}-\rho_{42}-\rho_{43}\right)
+{\Omega\over 2}e^{-i(\omega t+\varphi)}\left(-\rho_{24}-\rho_{34}+\rho_{48}\right)\eqno(V23)$$
$$(vN)_{45}=-\left(\omega_A-\omega_B-\omega_C\right)\rho_{45}-{\Omega\over 2}e^{i(\omega t+\varphi)}
\left(\rho_{85}-\rho_{41}\right)
+{\Omega\over 2}e^{-i(\omega t+\varphi)}\left(-\rho_{25}-\rho_{35}+\rho_{47}+\rho_{46}\right)\eqno(V24)$$
$$(vN)_{46}=-\left(\omega_A-\omega_B-j/2+j^{'}/2\right)\rho_{46}-{\Omega\over 2}e^{i(\omega t+\varphi)}
\left(\rho_{86}-\rho_{42}-\rho_{45}\right)
+{\Omega\over 2}e^{-i(\omega t+\varphi)}\left(-\rho_{26}-\rho_{36}+\rho_{48}\right)\eqno(V25)$$
$$(vN)_{47}=-\left(\omega_A-\omega_C\right)\rho_{47}-{\Omega\over 2}e^{i(\omega t+\varphi)}
\left(\rho_{87}-\rho_{43}-\rho_{45}\right)
+{\Omega\over 2}e^{-i(\omega t+\varphi)}\left(-\rho_{27}-\rho_{37}+\rho_{48}\right)\eqno(V26)$$
$$(vN)_{48}=-\left(\omega_A+j/2+j^{'}/2\right)\rho_{48}-{\Omega\over 2}e^{i(\omega t+\varphi)}
\left(\rho_{88}-\rho_{44}-\rho_{46}-\rho_{47}\right)
+{\Omega\over 2}e^{-i(\omega t+\varphi)}\left(-\rho_{28}-\rho_{38}\right)\eqno(V27)$$
$$(vN)_{55}=-{\Omega\over 2}e^{i(\omega t+\varphi)}\left(\rho_{65}+\rho_{75}-\rho_{51}\right)
+{\Omega\over 2}e^{-i(\omega t+\varphi)}\left(-\rho_{15}+\rho_{57}+\rho_{56}\right)\eqno(V28)$$
$$(vN)_{56}=-\left(\omega_C-j/2+j^{'}/2\right)\rho_{56}-{\Omega\over 2}e^{i(\omega t+\varphi)}
\left(\rho_{66}+\rho_{76}-\rho_{52}-\rho_{55}\right)
+{\Omega\over 2}e^{-i(\omega t+\varphi)}\left(-\rho_{16}+\rho_{58}\right)\eqno(V29)$$
$$(vN)_{57}=-\omega_B\rho_{57}-{\Omega\over 2}e^{i(\omega t+\varphi)}
\left(\rho_{67}+\rho_{77}-\rho_{53}-\rho_{55}\right)
+{\Omega\over 2}e^{-i(\omega t+\varphi)}\left(-\rho_{17}+\rho_{58}\right)\eqno(V30)$$
$$(vN)_{58}=-\left(\omega_B+\omega_C+j/2+j^{'}/2\right)\rho_{58}-{\Omega\over 2}e^{i(\omega t+\varphi)}
\left(\rho_{68}+\rho_{78}-\rho_{54}-\rho_{56}-\rho_{57}\right)
+{\Omega\over 2}e^{-i(\omega t+\varphi)}\left(-\rho_{18}\right)\eqno(V31)$$
$$(vN)_{66}=-{\Omega\over 2}e^{i(\omega t+\varphi)}\left(\rho_{86}-\rho_{62}-\rho_{65}\right)
+{\Omega\over 2}e^{-i(\omega t+\varphi)}\left(-\rho_{26}-\rho_{56}+\rho_{68}\right)\eqno(V32)$$
$$(vN)_{67}=-\left(\omega_B-\omega_C+j/2-j^{'}/2\right)\rho_{67}-{\Omega\over 2}e^{i(\omega t+\varphi)}
\left(\rho_{87}-\rho_{63}-\rho_{65}\right)
+{\Omega\over 2}e^{-i(\omega t+\varphi)}\left(-\rho_{27}-\rho_{57}+\rho_{68}\right)\eqno(V33)$$
$$(vN)_{68}=-\left(\omega_B+j\right)\rho_{68}-{\Omega\over 2}e^{i(\omega t+\varphi)}
\left(\rho_{88}-\rho_{64}-\rho_{66}-\rho_{67}\right)
+{\Omega\over 2}e^{-i(\omega t+\varphi)}\left(-\rho_{28}-\rho_{58}\right)\eqno(V34)$$
$$(vN)_{77}=-{\Omega\over 2}e^{i(\omega t+\varphi)}\left(\rho_{87}-\rho_{73}-\rho_{75}\right)
+{\Omega\over 2}e^{-i(\omega t+\varphi)}\left(-\rho_{37}-\rho_{57}+\rho_{78}\right)\eqno(V35)$$
$$(vN)_{78}=-\left(\omega_C+j/2+j^{'}/2\right)\rho_{78}-{\Omega\over 2}e^{i(\omega t+\varphi)}\left(\rho_{88}-\rho_{74}-\rho_{76}-\rho_{77}\right)
+{\Omega\over 2}e^{-i(\omega t+\varphi)}\left(-\rho_{38}-\rho_{58}\right)\eqno(V36)$$
$$(vN)_{88}=-{\Omega\over 2}e^{i(\omega t+\varphi)}\left(-\rho_{84}-\rho_{86}-\rho_{87}\right)
+{\Omega\over 2}e^{-i(\omega t+\varphi)}\left(-\rho_{48}-\rho_{68}-\rho_{78}\right)\eqno(V37)$$
\subsection*{Dissipation part.}

$$\mathcal{L}\rho_{11}=-\left(\gamma^{(1)}_A+\gamma^{(1)}_B+\gamma^{(1)}_C\right)\rho_{11}+
\gamma^{\dag (1)}_A\rho_{55}+\gamma^{\dag(1)}_B\rho_{33}+\gamma^{\dag(1)}_C\rho_{22},$$
$$\mathcal{L}\rho_{12}=-\left({\gamma^{(1)}_A+\gamma^{(2)}_A\over 2}+{\gamma^{(1)}_B+\gamma^{(2)}_B\over 2}+
{\gamma^{(1)}_C+\gamma^{\dag(2)}_C\over 2}\right)\rho_{12}+{\gamma^{\dag(1)}_A+\gamma^{\dag(2)}_A\over 2}\gamma^{(65)}_A(t)\rho_{56}+
{\gamma^{\dag(1)}_B+\gamma^{\dag(2)}_B\over 2}\gamma^{(43)}_B(t)\rho_{34}$$
$$\mathcal{L}\rho_{13}=-\left({\gamma^{(1)}_A+\gamma^{(3)}_A\over 2}+{\gamma^{(1)}_B+\gamma^{\dag (3)}_B\over 2}+
{\gamma^{(1)}_C+\gamma^{(3)}_C\over 2}\right)\rho_{13}+{\gamma^{\dag(1)}_A+\gamma^{\dag(3)}_A\over 2}\gamma^{(75)}_A(t)\rho_{57}+
{\gamma^{\dag(1)}_C+\gamma^{\dag(3)}_C\over 2}\gamma^{(42)}_C(t)\rho_{24}$$
$$\mathcal{L}\rho_{14}=-\left({\gamma^{(1)}_A+\gamma^{(4)}_A\over 2}+{\gamma^{(1)}_B+\gamma^{\dag(4)}_B\over 2}+
{\gamma^{(1)}_C+\gamma^{\dag(4)}_C\over 2}\right)\rho_{14}+{\gamma^{\dag(1)}_A+\gamma^{\dag(4)}_A\over 2}\gamma^{(85)}_A(t)\rho_{58}$$
$$\mathcal{L}\rho_{15}=-\left({\gamma^{(1)}_A+\gamma^{\dag(5)}_A\over 2}+{\gamma^{(1)}_B+\gamma^{(5)}_B\over 2}+
{\gamma^{(1)}_C+\gamma^{(5)}_C\over 2}\right)\rho_{15}+{\gamma^{\dag(1)}_B+\gamma^{\dag(5)}_B\over 2}\gamma^{(73)}_B(t)\rho_{37}+
{\gamma^{\dag(1)}_C+\gamma^{\dag(6)}_C\over 2}\gamma^{(62)}_C(t)\rho_{26}$$
$$\mathcal{L}\rho_{16}=-\left({\gamma^{(1)}_A+\gamma^{\dag(6)}_A\over 2}+{\gamma^{(1)}_B+\gamma^{(6)}_B\over 2}+
{\gamma^{(1)}_C+\gamma^{\dag(6)}_C\over 2}\right)\rho_{16}+{\gamma^{\dag(1)}_B+\gamma^{\dag(6)}_B\over 2}\gamma^{(83)}_B(t)\rho_{38}$$
$$\mathcal{L}\rho_{17}=-\left({\gamma^{(1)}_A+\gamma^{\dag(7)}_A\over 2}+{\gamma^{(1)}_B+\gamma^{\dag(7)}_B\over 2}+
{\gamma^{(1)}_C+\gamma^{(7)}_C\over 2}\right)\rho_{17}+{\gamma^{\dag(1)}_C+\gamma^{\dag(7)}_C\over 2}\gamma^{(82)}_C(t)\rho_{28}$$
$$\mathcal{L}\rho_{18}=-\left({\gamma^{(1)}_A+\gamma^{\dag(8)}_A\over 2}+{\gamma^{(1)}_B+\gamma^{\dag(8)}_B\over 2}+
{\gamma^{(1)}_C+\gamma^{\dag(8)}_C\over 2}\right)\rho_{18}$$
$$\mathcal{L}\rho_{22}=-\left(\gamma^{(2)}_A+\gamma^{(2)}_B+\gamma^{\dag(2)}_C\right)\rho_{22}
+\gamma^{\dag(2)}_A\rho_{66}+\gamma^{\dag(2)}_B\rho_{44}+\gamma^{(2)}_C\rho_{11}$$
$$\mathcal{L}\rho_{23}=-\left({\gamma^{(2)}_A+\gamma^{(3)}_A\over 2}+{\gamma^{(2)}_B+\gamma^{\dag(3)}_B\over 2}+
{\gamma^{(2)}_C+\gamma^{\dag(3)}_C\over 2}\right)\rho_{23}+{\gamma^{\dag(2)}_A+\gamma^{\dag(3)}_A\over 2}\gamma^{(76)}_A(t)\rho_{67}$$
$$\mathcal{L}\rho_{24}=-\left({\gamma^{(2)}_A+\gamma^{(4)}_A\over 2}+{\gamma^{(2)}_B+\gamma^{\dag(4)}_B\over 2}+
{\gamma^{\dag(2)}_C+\gamma^{\dag(4)}_C\over 2}\right)\rho_{24}+{\gamma^{\dag(2)}_A+\gamma^{\dag(4)}_A\over 2}\gamma^{(86)}_A(t)\rho_{68}+
{\gamma^{\dag(2)}_C+\gamma^{(4)}_C\over 2}\gamma^{(13)}_C(t)\rho_{13}$$
$$\mathcal{L}\rho_{25}=-\left({\gamma^{(2)}_A+\gamma^{\dag(5)}_A\over 2}+{\gamma^{(2)}_B+\gamma^{(5)}_B\over 2}+
{\gamma^{\dag(2)}_C+\gamma^{(5)}_C\over 2}\right)\rho_{25}+{\gamma^{\dag(2)}_B+\gamma^{\dag(5)}_B\over 2}\gamma^{(74)}_B(t)\rho_{47}$$
$$\mathcal{L}\rho_{26}=-\left({\gamma^{(2)}_A+\gamma^{\dag(6)}_A\over 2}+{\gamma^{(2)}_B+\gamma^{(6)}_B\over 2}+
{\gamma^{\dag(2)}_C+\gamma^{\dag(6)}_C\over 2}\right)\rho_{26}+{\gamma^{\dag(2)}_B+\gamma^{\dag(6)}_B\over 2}\gamma^{(84)}_B(t)\rho_{48}+
{\gamma^{(2)}_C+\gamma^{(6)}_C\over 2}\gamma^{(15)}_C(t)\rho_{15}$$
$$\mathcal{L}\rho_{27}=-\left({\gamma^{(2)}_A+\gamma^{\dag(7)}_A\over 2}+{\gamma^{(2)}_B+\gamma^{\dag(7)}_B\over 2}+
{\gamma^{\dag(2)}_C+\gamma^{(7)}_C\over 2}\right)\rho_{27}$$
$$\mathcal{L}\rho_{28}=-\left({\gamma^{(2)}_A+\gamma^{\dag(8)}_A\over 2}+{\gamma^{(2)}_B+\gamma^{\dag(8)}_B\over 2}+
{\gamma^{\dag(2)}_C+\gamma^{\dag(8)}_C\over 2}\right)\rho_{28}+{\gamma^{(2)}_C+\gamma^{(8)}_C\over 2}\gamma^{(17)}_A(t)\rho_{17}$$
$$\mathcal{L}\rho_{33}=-\left(\gamma^{(3)}_A+\gamma^{\dag(3)}_B+
\gamma^{(3)}_C\right)\rho_{33}+\gamma^{\dag(3)}_A\rho_{77}+\gamma^{(3)}_B\rho_{11}+
\gamma^{\dag(3)}_C\rho_{44}$$
$$\mathcal{L}\rho_{34}=-\left({\gamma^{(3)}_A+\gamma^{(4)}_A\over 2}+{\gamma^{\dag(3)}_B+\gamma^{\dag(4)}_B\over 2}+
{\gamma^{(3)}_C+\gamma^{\dag(4)}_C\over 2}\right)\rho_{34}+{\gamma^{\dag(3)}_A+\gamma^{\dag(4)}_A\over 2}\gamma^{(87)}_A(t)\rho_{78}+
{\gamma^{\dag(3)}_B+\gamma^{\dag(4)}_A\over 2}\gamma^{(12)}_B(t)\rho_{12}$$
$$\mathcal{L}\rho_{35}=-\left({\gamma^{(3)}_A+\gamma^{\dag(5)}_A\over 2}+{\gamma^{\dag(3)}_B+\gamma^{(5)}_B\over 2}+
{\gamma^{(3)}_C+\gamma^{(5)}_C\over 2}\right)\rho_{35}+{\gamma^{\dag(3)}_C+\gamma^{\dag(5)}_C\over 2}\gamma^{(64)}_A(t)\rho_{46}$$
$$\mathcal{L}\rho_{36}=-\left({\gamma^{(3)}_A+\gamma^{\dag(6)}_A\over 2}+{\gamma^{\dag(3)}_B+\gamma^{(6)}_B\over 2}+
{\gamma^{(3)}_C+\gamma^{\dag(6)}_C\over 2}\right)\rho_{36}$$
$$\mathcal{L}\rho_{37}=-\left({\gamma^{(3)}_A+\gamma^{\dag(7)}_A\over 2}+{\gamma^{\dag(3)}_B+\gamma^{\dag(7)}_B\over 2}+
{\gamma^{(3)}_C+\gamma^{(7)}_C\over 2}\right)\rho_{37}+{\gamma^{(3)}_B+\gamma^{(7)}_B\over 2}\gamma^{(15)}_B(t)\rho_{15}+
{\gamma^{\dag(3)}_C+\gamma^{\dag(7)}_C\over 2}\gamma^{(84)}_C(t)\rho_{48}$$
$$\mathcal{L}\rho_{38}=-\left({\gamma^{(3)}_A+\gamma^{\dag(8)}_A\over 2}+{\gamma^{\dag(3)}_B+\gamma^{\dag(8)}_B\over 2}+
{\gamma^{(3)}_C+\gamma^{\dag(8)}_C\over 2}\right)\rho_{38}+{\gamma^{\dag(3)}_B+\gamma^{\dag(8)}_B\over 2}\gamma^{(16)}_B(t)\rho_{16}$$
$$\mathcal{L}\rho_{44}=-\left(\gamma^{(4)}_A+\gamma^{\dag(4)}_B+\gamma^{\dag(4)}_C\right)\rho_{44}+\gamma^{\dag(4)}_A\rho_{88}
+\gamma^{(4)}_B\rho_{22}+\gamma^{(4)}_C\rho_{33}$$
$$\mathcal{L}\rho_{45}=-\left({\gamma^{(4)}_A+\gamma^{\dag(5)}_A\over 2}+{\gamma^{\dag(4)}_B+\gamma^{(5)}_B\over 2}+
{\gamma^{\dag(4)}_C+\gamma^{(5)}_C\over 2}\right)\rho_{45}$$
$$\mathcal{L}\rho_{46}=-\left({\gamma^{(4)}_A+\gamma^{\dag(6)}_A\over 2}+{\gamma^{\dag(4)}_B+\gamma^{(6)}_B\over 2}+
{\gamma^{\dag(4)}_C+\gamma^{\dag(6)}_C\over 2}\right)\rho_{46}+{\gamma^{(4)}_C+\gamma^{(6)}_C\over 2}\gamma^{(35)}_C(t)\rho_{35}$$
$$\mathcal{L}\rho_{47}=-\left({\gamma^{(4)}_A+\gamma^{\dag(7)}_A\over 2}+{\gamma^{\dag(4)}_B+\gamma^{\dag(7)}_B\over 2}+
{\gamma^{\dag(4)}_C+\gamma^{(7)}_C\over 2}\right)\rho_{47}+{\gamma^{(4)}_B+\gamma^{(7)}_B\over 2}\gamma^{(25)}_B(t)\rho_{25}$$
$$\mathcal{L}\rho_{48}=-\left({\gamma^{(4)}_A+\gamma^{\dag(8)}_A\over 2}+{\gamma^{\dag(4)}_B+\gamma^{\dag(8)}_B\over 2}+
{\gamma^{\dag(4)}_C+\gamma^{\dag(8)}_C\over 2}\right)\rho_{48}+{\gamma^{(4)}_B+\gamma^{(8)}_B\over 2}\gamma^{(26)}_B(t)\rho_{26}+
{\gamma^{(4)}_C+\gamma^{(8)}_C\over 2}\gamma^{(37)}_C(t)\rho_{37}$$
$$\mathcal{L}\rho_{55}=-\left(\gamma^{\dag(5)}_A+\gamma^{(5)}_B+\gamma^{(5)}_C\right)\rho_{55}+\gamma^{(5)}_A\rho_{11}
+\gamma^{\dag(5)}_B\rho_{77}+\gamma^{\dag(5)}_C\rho_{66}$$
$$\mathcal{L}\rho_{56}=-\left({\gamma^{\dag(5)}_A+\gamma^{\dag(6)}_A\over 2}+{\gamma^{(5)}_B+\gamma^{(6)}_B\over 2}+
{\gamma^{(5)}_C+\gamma^{\dag(6)}_C\over 2}\right)\rho_{56}+{\gamma^{(5)}_A+\gamma^{(6)}_A\over 2}\gamma^{(12)}_A(t)\rho_{12}+
{\gamma^{\dag(5)}_B+\gamma^{\dag(6)}_B\over 2}\gamma^{(87)}_B(t)\rho_{78}$$
$$\mathcal{L}\rho_{57}=-\left({\gamma^{\dag(5)}_A+\gamma^{\dag(7)}_A\over 2}+{\gamma^{\dag(5)}_B+\gamma^{\dag(7)}_B\over 2}+
{\gamma^{(5)}_C+\gamma^{(7)}_C\over 2}\right)\rho_{57}+{\gamma^{(5)}_A+\gamma^{(7)}_A\over 2}\gamma^{(13)}_A(t)\rho_{13}+
{\gamma^{\dag(5)}_C+\gamma^{\dag(7)}_C\over 2}\gamma^{(86)}_C(t)\rho_{68}$$
$$\mathcal{L}\rho_{58}=-\left({\gamma^{\dag(5)}_A+\gamma^{\dag(8)}_A\over 2}+{\gamma^{(5)}_B+\gamma^{\dag(8)}_B\over 2}+
{\gamma^{(5)}_C+\gamma^{\dag(8)}_C\over 2}\right)\rho_{58}+{\gamma^{(5)}_A+\gamma^{(8)}_A\over 2}\gamma^{(14)}_A(t)\rho_{14}$$
$$\mathcal{L}\rho_{66}=-\left(\gamma^{\dag(6)}_A+\gamma^{(6)}_B+\gamma^{\dag(6)}_C\right)\rho_{66}+
\gamma^{(6)}_A\rho_{22}+\gamma^{\dag(6)}_B\rho_{88}+\gamma^{(6)}_C\rho_{55}$$
$$\mathcal{L}\rho_{67}=-\left({\gamma^{\dag(6)}_A+\gamma^{\dag(7)}_A\over 2}+{\gamma^{(6)}_B+\gamma^{\dag(7)}_B\over 2}+
{\gamma^{(6)}_C+\gamma^{\dag(7)}_C\over 2}\right)\rho_{67}+{\gamma^{(6)}_A+\gamma^{(7)}_A\over 2}\gamma^{(23)}_A(t)\rho_{23}$$
$$\mathcal{L}\rho_{68}=-\left({\gamma^{\dag(6)}_A+\gamma^{\dag(8)}_A\over 2}+{\gamma^{\dag(6)}_B+\gamma^{\dag(8)}_B\over 2}+
{\gamma^{\dag(6)}_C+\gamma^{\dag(8)}_C\over 2}\right)\rho_{68}+{\gamma^{(6)}_A+\gamma^{(8)}_A\over 2}\gamma^{(24)}_A(t)\rho_{24}+
{\gamma^{(6)}_C+\gamma^{(8)}_C\over 2}\gamma^{(57)}_C(t)\rho_{57}$$
$$\mathcal{L}\rho_{77}=-\left(\gamma^{\dag(7)}_A+\gamma^{\dag(7)}_B+\gamma^{(7)}_C\right)\rho_{77}+\gamma^{(7)}_A\rho_{33}
+\gamma^{(7)}_B\rho_{55}+\gamma^{\dag(7)}_C\rho_{88}$$
$$\mathcal{L}\rho_{78}=-\left({\gamma^{\dag(7)}_A+\gamma^{\dag(8)}_A\over 2}+{\gamma^{\dag(7)}_B+\gamma^{\dag(8)}_B\over 2}+
{\gamma^{(3)}_C+\gamma^{\dag(8)}_C\over 2}\right)\rho_{78}+{\gamma^{(7)}_A+\gamma^{(8)}_A\over 2}\gamma^{(34)}_A(t)\rho_{34}+
{\gamma^{(7)}_B+\gamma^{(8)}_B\over 2}\gamma^{(56)}_B(t)\rho_{56}$$
$$\mathcal{L}\rho_{88}=-\left(\gamma^{\dag(8)}_A+\gamma^{\dag(8)}_B+\gamma^{\dag(8)}_C\right)\rho_{88}+\gamma^{(8)}_A\rho_{44}
+\gamma^{(8)}_B\rho_{66}+\gamma^{(8)}_C\rho_{77}$$

\subsection*{Eigenvalues of the $\Omega_k$ operator.}
The eigenvalues equation is written as $\Omega_A|i\rangle= \Omega_A^{(i)}|i\rangle $, 
for a three nuclear spins $|ABC\rangle$.  The basis is taken in decimal notation, like $|1\rangle=|000\rangle$, $|2\rangle=|001\rangle$,
and so on.
$$\Omega^{(1)}_A=\Omega^{(5)}_A=\omega_A-J/2-J'/2, \quad \Omega^{(1)}_B=\Omega^{(3)}_B=\omega_B-J, 
\quad \Omega^{(1)}_C=\Omega^{(2)}_C=\omega_C-J/2-J'/2,\eqno{(A1)}$$ 
$$\Omega^{(2)}_A=\Omega^{(6)}_A=\omega_A-J/2+J'/2, \quad \Omega^{(2)}_B=\Omega^{(4)}_B=\omega_B,
\quad \Omega^{(3)}_C=\Omega^{(4)}_C=\omega_C+J/2-J'/2,\eqno{(A2)}$$ 
$$\Omega^{(3)}_A=\Omega^{(7)}_A=\omega_A+J/2-J'/2, \quad \Omega^{(5)}_B=\Omega^{(7)}_B=\omega_B,
\quad \Omega^{(5)}_C=\Omega^{(6)}_C=\omega_C-J/2+J'/2,\eqno{(A3)}$$ 
$$\Omega^{(4)}_A=\Omega^{(8)}_A=\omega_A+J/2+J'/2, \quad \Omega^{(6)}_B=\Omega^{(8)}_B=\omega_B+J,
\quad \Omega^{(7)}_C=\Omega^{(8)}_C=\omega_C+J/2+J'/2, \eqno{(A4)}$$

\end{document}